\documentclass{article}

\usepackage{PRIMEarxiv}

\usepackage[utf8]{inputenc} % allow utf-8 input
\usepackage[T1]{fontenc}    % use 8-bit T1 fonts
\usepackage{hyperref}       % hyperlinks
\usepackage{url}            % simple URL typesetting
\usepackage{booktabs}       % professional-quality tables
\usepackage{amsfonts}       % blackboard math symbols
\usepackage{nicefrac}       % compact symbols for 1/2, etc.
\usepackage{microtype}      % microtypography
\usepackage{lipsum}
\usepackage{fancyhdr}       % header
\usepackage{graphicx}       % graphics
\graphicspath{{media/}}     % organize your images and other figures under media/ folder
\usepackage{amssymb}
\usepackage{csquotes}
\usepackage{textgreek}
\usepackage{float}
\usepackage{siunitx}
\usepackage{physics}
\usepackage{multirow}

\usepackage{caption}
\usepackage{subcaption}
\usepackage{xcolor}

\usepackage{booktabs}

\usepackage{empheq, etoolbox}
\newcommand{\eref}[1]{Eq.~\eqref{#1}}

\newcommand{\fref}[1]{Fig.~\ref{#1}}
\newcommand{\Fref}[1]{Figure~\ref{#1}}

\newcommand{\sref}[1]{section~\ref{#1}}
\newcommand{\Sref}[1]{Section~\ref{#1}}

\newcommand{\Ng}{{N\textsubscript{g}}}
\newcommand{\Ns}{{N\textsubscript{s}}}

\newcommand{\Sg}{{S\textsubscript{g}}}
\newcommand{\So}{{S\textsubscript{0}}}

\newcommand{\Rg}{$R_{g}$}
\newcommand{\Ro}{$R_{0}$}

\newcommand{\Rlim}{$R_{lim}$}

\title{Particle pinning during grain growth - A new analytical model for predicting the mean limiting grain size but also grain size heterogeneity in a 2D polycrystalline context 
}

\author{
  Madeleine Bignon, Marc Bernacki \\
  Mines Paris, PSL University\\
  Centre for material forming (CEMEF), UMR CNRS\\
  06904 Sophia Antipolis, France\\
  \texttt{\{madeleine.bignon, marc.bernacki\}@minesparis.psl.eu} \\
}

\begin{document}
\maketitle

\begin{abstract}
This study proposes a new analytical model for grain boundary pinning by second phase particles in two-dimensional polycrystals. This approach not only considers how particles impede grain growth, but also elucidates their role in preventing grain disappearance, thereby leading to stabilised microstructures characterised by heterogeneous grain size distribution comprising a mixture of small and large grains. By quantifying the number of particles intercepted by grain boundaries during grain growth or shrinkage, we are able to calculate the respective sizes and fractions of large and small grains. Furthermore, we identify ranges of particle surface fractions and particle sizes that maximise the heterogeneity in grain size. Additionally, we demonstrate the significant influence of initial grain size on the limiting grain size in pinned microstructures. Our analytical model's results are compared with those obtained from full-field level-set simulations conducted in this study and from phase-field calculations reported in the literature, revealing very good agreement. Finally, the differences between the proposed model and existing ones in the literature are discussed.
\end{abstract}

% keywords can be removed
\keywords{Smith-Zener pinning \and Grain growth \and Microstructure evolution \and metallurgy \and modelling \and full-field simulations}

\section{Introduction}

The control of grain size in polycrystalline materials, such as metallic alloys, ceramics, minerals, or composites, is an important issue. Second phase particles are commonly used to limit capillarity-driven grain growth in materials such as steels \cite{Karasev2006}, nickel alloys, aluminium alloys, magnesium alloys \cite{Robson2011}, or alumina-zirconia composites \cite{Alexander1994}. This phenomenon is known as particle pinning, or Smith-Zener pinning, and usually results in the grains reaching a stagnated size $R_{\rm{lim}}$. Predicting the effects of particles on grain growth is essential for controlling microstructure and thus material properties.

Particle pinning occurs when a grain boundary meets a particle, reducing the total surface occupied by grain boundaries and consequently lowering the system's free energy. This phenomenon was first rationalised by Smith and Zener who proposed a model for stagnated grain size in particle containing materials  \cite{Smith1948}. In their original treatment, the limiting grain size is calculated as the critical size at which the pressure $P_{\gamma}$, driven by the reduction in surface energy, is balanced by an equivalent pressure $P_{z}$ exerted by all particles in contact with the boundary \cite{Smith1948, Nes1985}. The limiting grain size $R_{\rm{lim}}$ is determined such that $P_{z}=P_{\gamma}$, and the grain boundaries are thought to detach from particles when the pressure induced by capillarity exceeds that exerted by particles ($|P_{\gamma}|>|P_{z}|$).

Since Smith and Zener's initial treatment, particle pinning modelling has attracted significant attention in the material science community, leading to the development of increasingly sophisticated analytical models.  Most of these treatments converge on how to treat the interaction between a single segment of grain boundary and a single particle, as originally suggested by Smith and Zener \cite{Smith1948} . The problem has been much more debated when it comes to dealing with a polycrystal containing an array of particles \cite{Miodownik2000, Harun2006}.  

Various modelling strategies have been proposed in the literature to enhance the Smith-Zener model, many of which are summarised in \cite{Nes1985, Manohar1998}. Some authors modified the calculation of $P_{z}$  by considering the attractive effect of particles on boundaries \cite{Louat1982, Hunderi1989}. Others demonstrated that the pinning pressure should depend on the relative position of the particle centres with respect to the boundary and proposed a corrected expression of $P_{z}$ \cite{Louat1982, Hunderi1989, Hellman1975, Worner1987}. Other modifications involve models where the stagnated grain size is influenced by the particle location \cite{Nishizawa1997} or size distribution  \cite{Flowers1979, Haroun1980}. Some researchers also attempted to develop more realistic expressions for the grain boundary driving pressure \cite{Haroun1980}.  

Most of these models, although starting from varying hypotheses, reach the conclusion that the limiting grain size follows a law of the form $R_{\rm{lim}}=\beta r_{p}/f_{v}^{m}$, where $f_{v}$ and $r_{p}$ are the particle volume fraction and mean radius, respectively, and $m$ and $\beta$ are model parameters \cite{Manohar1998}. In this formulation, $R_{\rm{lim}}$ is the radius of equivalent surface to that of the stagnated grain.  Manohar's review demonstrated that no unique pair of values for $m$ and $\beta$ can accurately predict the stagnated grain size across the full range of particle volume fractions found in engineering materials \cite{Manohar1998}. The parameters $\beta$ and $m$ are thus usually fitted to experimental data over a restricted range of particle volume fraction. As a result, there is room for improvement to predict particle pinning in a more reliable way and without the need for \textit{ad-hoc} experimental data.

The long-term goal of this study is to propose a simple analytical model, free of empirical parameter, and capable of accurately predicting stagnated grain size across a wide range of particle volume fractions. For this purpose, this work presents, as a first step, a model predicting limiting grain sizes in two-dimensional microstructures. The main difference between this model and existing ones from literature lies in the way to address unpinning. The present model considers that, in the absence of difference in plastic stored energy between adjacent grains, grain boundaries are not able to break free from particles. This results in particles not only inhibiting grain growth, but also limiting grain shrinkage. The model developed in this work shows how the stagnated grain sizes in two-dimensional microstructures can be simply predicted knowing the initial grain size, the mean free path between particles, and their surface density. The proposed model does not result on a relation of the type $R_{\rm{lim}}=\beta r_{p}/f_{v}^{m}$ . 

The physical situation described by the model developed here corresponds to a fibre-type microstructure \cite{Nishizawa1997}. Unfortunately, experimental data regarding particle-controlled grain growth in such a microstructure is not available as far as we are aware. As a result, our model cannot directly be validated using experimental data. On the other hand, mesoscale two-dimensional full-field simulations have long been acknowledged as a precious tool to challenge particle pinning analytical models. Examples of such simulations include Monte Carlo simulations \cite{Gao1997, Srolovitz1984, Hazzledine1990}, phase field models \cite{Moelans2006, Manna2023}, front tracking models \cite{Weygand1999, Harun2006} or level-set formulation \cite{dervieux1980, Osher1988, Osher2001, Osher2004, Agnoli2014}. 

In this work, mesoscale full-field simulations have been performed, using an already existing level-set numerical framework described in \cite{Agnoli2014}. These simulations offer an advantageous means of validation as they operate independently of many of the assumptions inherent in the analytical model. Additionally, simulations provide the flexibility to explore a wide range of particle surface fractions and initial grain sizes while circumventing complicating factors such as particle formation, dissolution or coarsening, as well as recrystallisation, which would otherwise complicate the validation process.

\section{Analytical model for limiting grain size} \label{section:analytic_model}
The problem considered in this work is that of a two-dimensional polycrystal, where grain boundaries interact with circular stable second phase particles of uniform size, randomly distributed in space, that show no tendency to grow, dissolve or coarsen. We consider a microstructure free of deformation energy, where there is no difference in free energy between the grains ahead and behind the migrating boundaries. Thus, the boundaries only migrate under the effect of their own surface energy $\gamma$, which is assumed homogeneous.
\subsection{Behaviour of a grain boundary meeting particles}

\begin{figure}[H]

\begin{subfigure}{0.4\textwidth}

\includegraphics[trim={6cm 20cm 4.5cm 5cm},clip, scale=0.55]{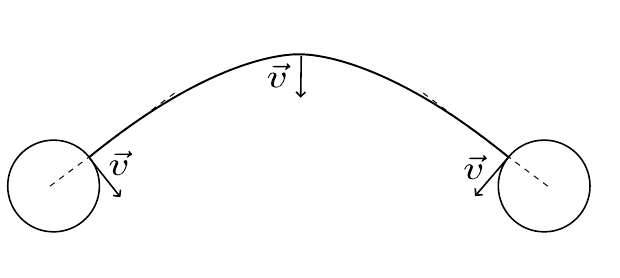} 
\caption{}
\label{subfig:two_particles_initial}
\end{subfigure}
\hfill
\begin{subfigure}{0.4\textwidth}
\center
\includegraphics[trim={6cm 20cm 4.5cm 5cm},clip, scale=0.55]{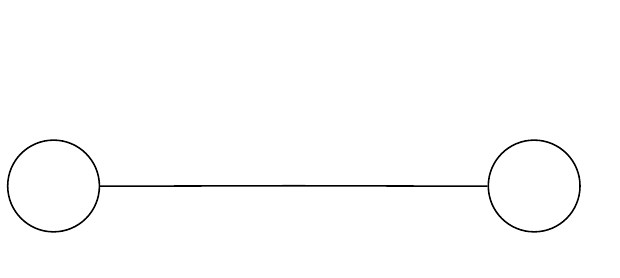} 
\caption{}
\label{subfig:two_particles_equilibrium}
\end{subfigure}
\begin{subfigure}{0.4\textwidth}

\includegraphics[trim={6cm 20cm 4.5cm 5cm},clip, scale=0.55]{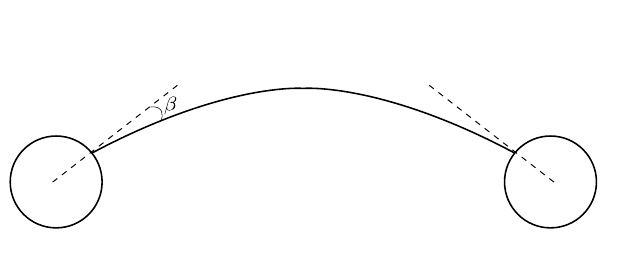} 
\caption{}
\label{subfig:two_particles_initial_incoherent}
\end{subfigure}
\hfill
\begin{subfigure}{0.4\textwidth}
\includegraphics[trim={6cm 20cm 4.5cm 5cm},clip, scale=0.55]{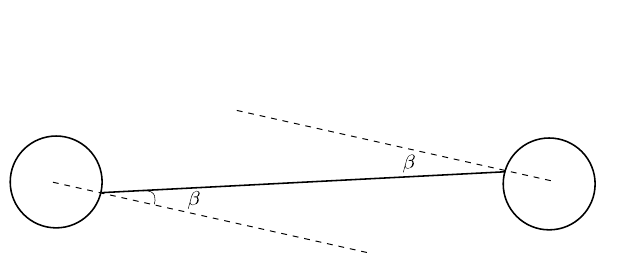} 
\caption{}
\label{subfig:two_particles_final_incoherent}
\end{subfigure}

\caption{(a) Grain boundary in contact with two incoherent particles. The curvature pushes the boundary towards its center of curvature. (b) Stable position of the grain boundary. Once the boundary forms a straight line between the particles, the configuration is stable and it no longer moves. (c) Grain boundary in contact with two particles, with non-zero angle between particle and boundary. The curvature pushes the boundary towards its center of curvature. (d) Stable position of the grain boundary. Once the boundary forms a straight line between the particles, the configuration is stable and it no longer moves.}
\label{fig:two_particles}
\end{figure}

We first consider the behaviour of a grain boundary, that is migrating under its own curvature and meets two particles. This situation is shown in \fref{subfig:two_particles_initial}. The angle formed between the boundary and the normal to the particle at the junction between them is determined by the coherency of the particle \cite{Ashby1968}, and it is assumed hereafter that the boundary meets the particle at right angle, which corresponds to incoherent particles \cite{Nes1985}. This assumption is not expected to induce any loss of generality for the model that follows, as will be discussed later.  

A boundary segment, as illustrated in \fref{subfig:two_particles_initial}, is subject to a driving pressure that pulls it towards its center of curvature at a velocity of $||\vec{v}|| = M \cdot \gamma  \cdot \kappa$, where $\kappa$ represents the local curvature, and $M$ is the grain boundary mobility. This motion aims to minimise the overall grain boundary energy, leading the boundary to adopt a configuration that minimises its length, that is a straight line between the two particles (\fref{subfig:two_particles_equilibrium}). Once the boundary assumes this straight-line configuration, it ceases to move. Any deviation from this stable configuration, shown in \fref{subfig:two_particles_equilibrium}, would result in an increase in total length, consequently raising the system's free energy. As a result, in a two-dimensional microstructure lacking free energy disparity between neighbouring grains, a grain boundary cannot disengage from precipitates. This holds true for coherent and semi-coherent particles that do not intersect the boundary normally, as depicted in \fref{subfig:two_particles_initial_incoherent} and \fref{subfig:two_particles_final_incoherent}.

\begin{figure}[H]
\center
\includegraphics[trim={2cm 15cm 5cm 2cm},clip, scale=0.4]{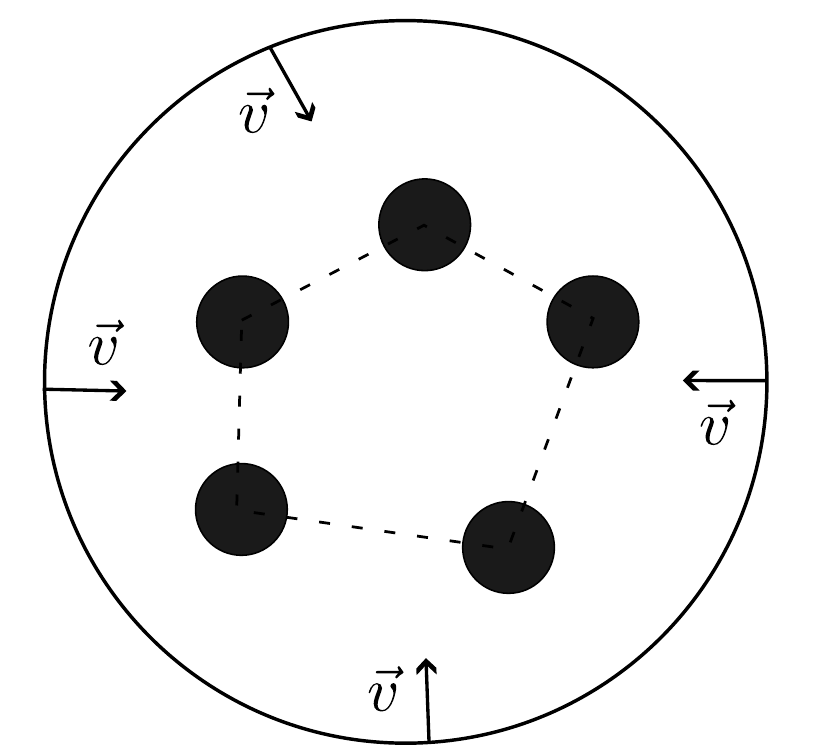} 
\caption{Circular grain migrating toward its center of curvature, with particles inside the grain. The plain line shows the initial configuration and the dotted line shows the pinned or stabilised configuration. }
\label{fig:circular_grain}
\end{figure}

Considering this, it becomes evident that an initially circular grain embedded in an homogeneous medium should migrate towards its center of curvature until it adopts a configuration where it forms straight lines between particles, as depicted in \fref{fig:circular_grain}. This aspect has been pointed out by several authors before \cite{Nishizawa1997, Kim1999}. It is noteworthy that, once the boundaries of a grain align into straight lines between precipitates, the grain no longer experiences any driving pressure, irrespective of its size. This aspect will be discussed in more detail later.

\subsection{Calculation of the limiting grain sizes}

\begin{figure}[H]

\begin{subfigure}{0.4\textwidth}
\includegraphics[trim={3.5cm 13cm 4.5cm 3.5cm},clip, scale=0.6]{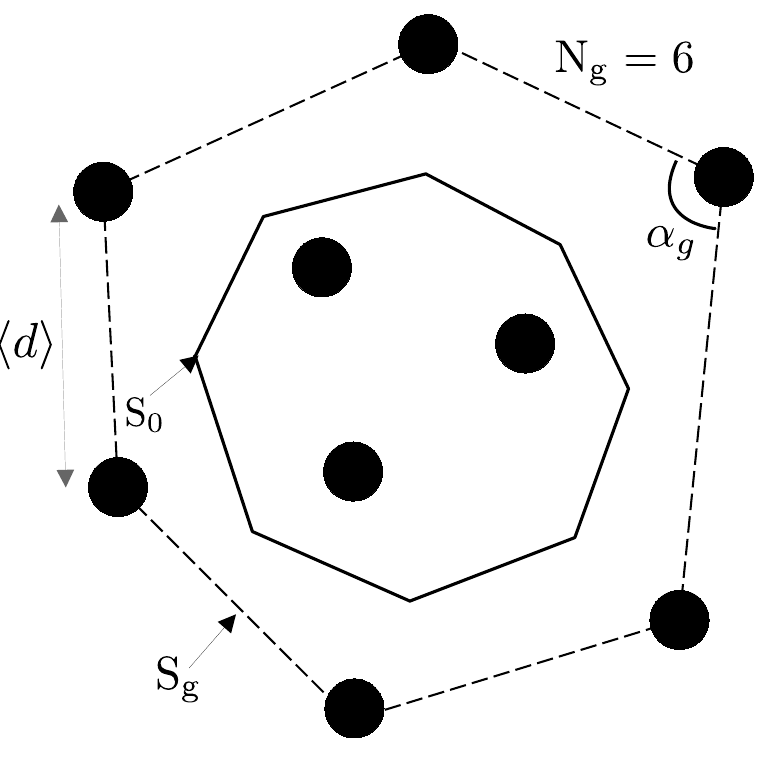} 
\caption{}
\label{subfig:growing_grain}
\end{subfigure}
\hfill
\begin{subfigure}{0.4\textwidth}
\includegraphics[trim={5cm 13cm 5cm 4.5cm},clip, scale=0.6]{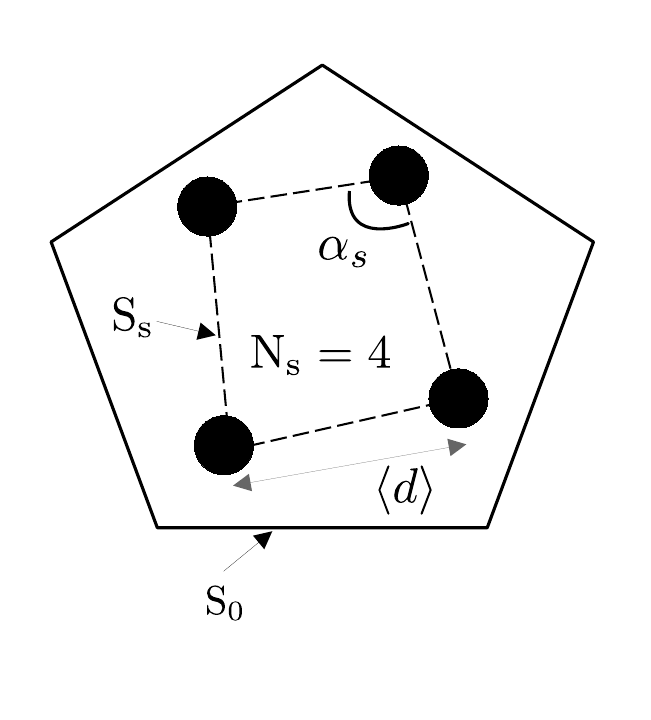} 
\caption{}
\label{subfig:shrinking_grain}
\end{subfigure}
\hfill

\caption{Schematic representation of (a) a growing grain; (b) a shrinking grain in the presence of particles. The plain lines show the initial grain boundary while the dotted line represents the grain boundary once the grain has reached its stable configuration.}
\label{fig:schematic_radius}
\end{figure}

When the boundary of a grain of initial surface S\textsubscript{0} moves (Fig.~\ref{fig:schematic_radius}), the grain can either grow (Fig.~\ref{subfig:growing_grain}) or shrink (Fig.~\ref{subfig:shrinking_grain}) . In the absence of second phase particles, there is no theoretical limit for shrinkage or growth; the largest grains keep growing until a metastable configuration is reached while smallest grains keep shrinking, and eventually disappear. 

In the case where second phase particles are present, however, the situation is different. As mentioned above, in two dimensions and in the absence of any driving pressure other than capillarity, grain boundaries do not break free from particles \cite{Kim1999}, and simply assume the shape that reduces the grain boundary length to its minimum. This leads to three possible evolutions for a grain of initial surface S\textsubscript{0}. (a) The grain grows until it reaches a stable configuration where all segments between the particles encountered during growth are straight lines (Fig.~\ref{subfig:growing_grain}). The grain then reaches an equilibrium surface S\textsubscript{g} and stops growing. (b) The grain shrinks and there are at least three particles inside the grain, as in Fig.~\ref{subfig:shrinking_grain}. It keeps shrinking until it reaches an equilibrium configuration where all segments between the particles encountered during shrinking are straight lines. The grain then reaches an equilibrium surface S\textsubscript{s} and stops shrinking. (c) The grain shrinks and there are less than three particles inside the initial grain of surface S\textsubscript{0}. The grain keeps shrinking and eventually disappears. 

Therefore, in the presence of particles, there should not be a single equilibrium grain size, as usually assumed in the literature, but two equilibrium grain sizes, namely $R_{g}$ and $R_{s}$, corresponding respectively to a growing grain limited in its growth by the presence of particles, and to a shrinking grain prevented from vanishing by the particles. The calculation of the two possible equilibrium grain sizes $R_{g}$ and $R_{s}$ is derived below. 

\subsubsection{Equilibrium size for a growing grain} \label{sec:equations_growing_grain}
Accepting that a grain boundary cannot break away from particles in the absence of stored energy, the number of particles N\textsubscript{g} intercepting a grain boundary when it has grown and eventually reached its equilibrium surface S\textsubscript{g} (\fref{fig:schematic_radius}) can be calculated by multiplying the area swept by the grain during its growth by the number of particles per surface unit $\rho_{s}$: 
\begin{equation} \label{eq:Ng}
\rm{N\textsubscript{g}} \cdot \frac{\alpha_{g}}{2 \pi}= \rho_{s}\cdot (S_{g}-S_{0})
\end{equation}
where $\alpha_{g}$ is the average angle between two segments of the boundary (Fig.~\ref{subfig:growing_grain}), that can be related to N\textsubscript{g} using the formula for the sum of the angles of a polygon:

\begin{equation}\label{eq:alpha_g}
\alpha_{g}=\frac{(\rm{\Ng-2})\pi}{\rm{\Ng}} 
\end{equation}

The factor $\frac{\alpha_{g}}{2 \pi}$ in \eref{eq:Ng} arises because only a fraction of each particle intercepting the boundary is inside the grain. In \eref{eq:Ng}, N\textsubscript{g} is unknown, but can be related to the number density of the particles. Indeed, if the grain is approximated by a regular polygon with N\textsubscript{g} sides of length $\left<d\right>$  as in \fref{fig:schematic_radius}, the surface of the grain is:
\begin{equation}
\rm{S\textsubscript{g}}=\frac{ \left<d\right>^{2}}{4}\cdot \rm{N\textsubscript{g}}\cdot cotan\left(\frac{\pi}{\rm{N\textsubscript{g}}}\right)
\end{equation}
where  $\left<d\right>$ can be eliminated by considering that, once the grain has reached its equilibrium size, the mean length of its sides $\left<d\right>$ is equal to the average free path between particles, and can thus be calculated as a function of the particle surface fraction $f_{s}$ and particle radius $r_{p}$ as:
\begin{equation} \label{eq:mean_free_path}
\left<d\right>=\frac{1}{\sqrt{\rho_{s}}}=\sqrt{\frac{\pi}{f_{s}}} \cdot r_{p}
\end{equation} 

When considering grain growth in presence of particles, it is usually more convenient to express the equilibrium configuration in terms of grain size rather than grain surface. The initial and equilibrium grain sizes \Ro~and \Rg~are defined here as the radii of the circles of equivalent surfaces (\So$=\pi\cdot \textrm{\Ro}^{2} $, \Sg$=\pi\cdot \textrm{\Rg}^{2} $). For convenience, we also define the normalised grain sizes as $r_{0}=\textrm{\Ro}/r_{p}$ and $r_{g}=\textrm{\Rg}/r_{p}$. Finally, a growing grain of initial normalised size $r_{0}$ stops growing when it reaches an equilibrium size $r_{g}$ such that - (Eq.~\ref{eq:Ng}~to~\ref{eq:mean_free_path}):

\begin{subequations} \label{eq:system}
  \begin{empheq}[left=\empheqlbrace]{align}   
      (r_{g}^{2}-r_{0}^{2})f_{s}=\dfrac{\rm{ \Ng}-2}{2} \label{subeq:system_growing_a}\\
      r_{g}^{2}=\dfrac{\rm{ \Ng}}{4f_{s}\tan(\dfrac{\pi}{\rm{\Ng}})}
   \end{empheq}
\end{subequations}

For all values of $r_{0}$ and $f_{s}$ considered here, the equation system (\ref{eq:system}) admits a unique solution for the couple (\Ng, $r_{g}$). 
\subsubsection{Equilibrium size for a shrinking grain}

If the shape of a grain of initial normalised size $r_{0}$ is such that this grain shrinks, and if there are more than two particles inside the grain, as shown in Fig.~\ref{subfig:shrinking_grain}, then the grain stops shrinking when it reaches an equilibrium normalised size $r_{s}$. At that point, the polygonal grain has \Ns~sides, and the values of \Ns~and $r_{s}$ can be calculated in a similar way as in \Sref{sec:equations_growing_grain}, giving: 
\begin{subequations} \label{eq:system_shrinking}
  \begin{empheq}[left=\empheqlbrace]{align}   
      (r_{0}^{2}-r_{s}^{2})f_{s}=\dfrac{\rm{ \Ns}-2}{2}\label{subeq:system_shrinking_a}\\
      r_{s}^{2}=\dfrac{\rm{\Ns}}{4f_{s}\tan(\dfrac{\pi}{\rm{\Ns}})}
   \end{empheq}
\end{subequations}
where the difference with the system of equation (\ref{eq:system}) lies on the fact that the grain has shrunk rather than grown, resulting in a difference on the left hand side of \eref{subeq:system_shrinking_a} compared to the left hand side of \eref{subeq:system_growing_a}.

\subsubsection{Equilibrium fractions of large and small grains}

Once the mean size of large and small grains are calculated, it is useful to calculate the average grain size \Rlim~in the pinned microstructure. Throughout this work, the average grain size is defined as: 

\begin{equation}
R_{lim}=\sqrt{\frac{\bar{\rm{S}}}{\pi}}=\sqrt{\frac{\sum\limits_{i=1}\limits^{n_{f}}{\rm{S_{i}}}}{\pi n_{f}}}
\end{equation}
where $n_{f}$ is the total number of grains in the final microstructure and $\rm{S_{i}}$ the surface of grain number $i$ (for $i=1..n_{p}$) in the pinned microstructure.

\begin{figure}[H]
\centering
\includegraphics[trim={5cm 19cm 1cm 4.5cm},clip, scale=1.2]{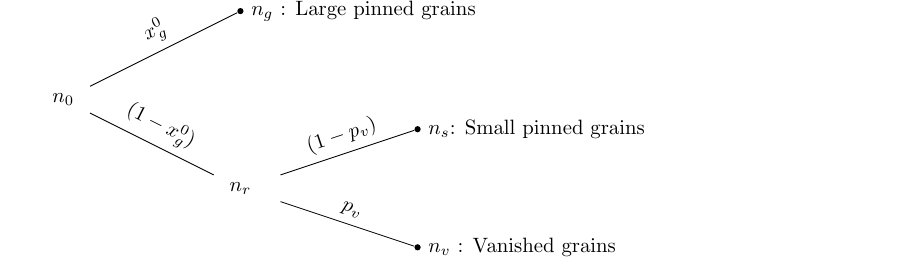} 
\caption{Probability tree for the evolution of an initial microstructure containing $n_{0}$ grains.  }
\label{fig:probability_tree}
\end{figure}

To compute the average grain size \Rlim~in the pinned microstructure, the respective fractions of large and small grains need to be estimated.  Considering an initial microstructure containing $n_{0}$ grains, these grains can be divided into $n_{g}$ grains that grow, and $n_{r}$ grains that contract. Among the latter, those that contain at least three particles inside the grain will be pinned ($n_{s}$), while the others ($n_{v}$) vanish. The situation is schematised in \fref{fig:probability_tree}. By definition, $n_{0}=n_{g}+n_{s}+n_{v}$ (and $n_{f}=n_{g}+n_{s}$), and it is convenient to define the fractions of grains of the initial microstructure that grow, vanish, or shrink and get pinned, respectively labelled $x^{0}_{g}$, $x^{0}_{v}$, and $x^{0}_{s}$, and verifying:
\begin{equation}\label{eq:sum_fractions}
x^{0}_{g}+x^{0}_{s}+x^{0}_{v}=1
\end{equation}
If the probability $p_{v}$ for a contracting grain to vanish is known, $x^{0}_{v}$ can be expressed as: 
\begin{equation}\label{eq:fractions}
x^{0}_{v}=(1-x^{0}_{g})\cdot p_{v}
\end{equation}

Considering that the total surface of the microstructure is the same at the beginning and at the end of the simulation, it is clear that $n_{0}\pi r_{0}^{2}= n_{s}\pi r_{s}^{2}+n_{g}\pi r_{g}^{2}$, and thus that:
\begin{equation} \label{eq:surface_conservation}
r_{0}^{2}= x_{s}^{0}r_{s}^{2}+x_{g}^{0}r_{g}^{2}
\end{equation}
Combining Equations (\ref{eq:sum_fractions}), (\ref{eq:fractions}) and (\ref{eq:surface_conservation}), the fraction of grains in the initial microstructure that have grown, shrunk without disappearing, and vanished, respectively, can be expressed as:

\begin{subequations} \label{eq:fraction_calculation}
  \begin{empheq}{align}   
     & x^{0}_{g}=\dfrac{r_{0}^{2}-(1-p_{v})r_{s}^{2}}{r_{g}^{2}-(1-p_{v})r_{s}^{2}}\\
     & x^{0}_{s}=(1-p_{v})(1-x_{g}^{0}) \\
    &  x^{0}_{v}=p_{v}(1-x^{0}_{g})
   \end{empheq}
\end{subequations}

\begin{figure}[H]
\centering
\includegraphics[trim={0cm 15cm 10cm 3.4cm},clip, scale=0.7]{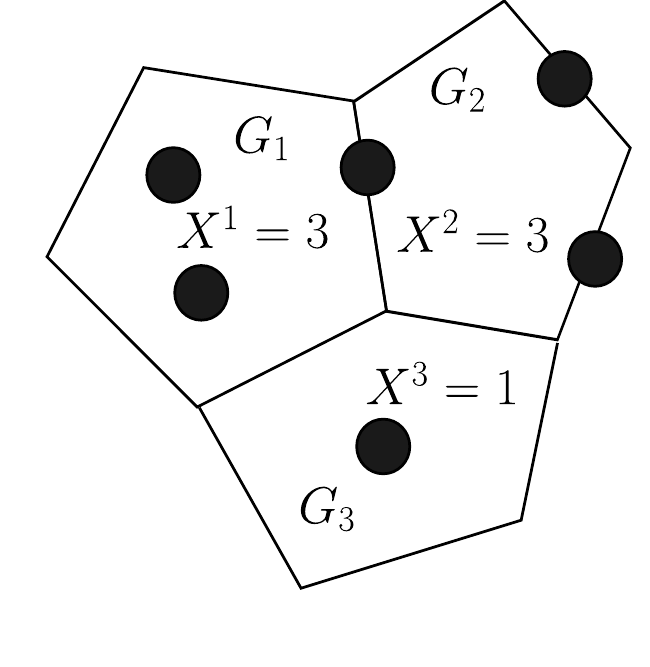} 
\caption{Illustration of the intersections between grain surface and particles in the initial microstructure. For each grain $G_{i}$, $X^{i}$ is the number of particles intersecting the grain. }
\label{fig:intersection_grain_particles}
\end{figure}

The probability $p_{v}$ for a contracting grain to vanish now needs to be calculated. As previously mentioned, whether a contracting grain $G_{i}$ can vanish or not depends on the number of particles $X^{i}$ that intersect with the surface of the grain $G_{i}$ in the initial microstructure. It should be noted that in this context, $X^{i}$ is necessarily an integer. In other words, every particle which intersects with the initial surface of the grain should be accounted for in $X^{i}$, irrespective of whether the particle touches the interface of the grain or is embedded in the grain. This is illustrated in \fref{fig:intersection_grain_particles}.

Consider an initial microstructure of $n_{0}$ grains of radius $R_{0}$, populated with $N_{p}$ randomly distributed particles of radius $r_{p}$ and surface fraction $f_{s}$.  The probability $p(M_{j} \cap G_{i})$ that a particle $M_{j}$ and a grain $G_{i}$ intersect is:

\begin{equation}
 p(M_{j} \cap G_{i})=\frac{\pi(R_{0}+r_{p})^{2}}{n_{0}\pi R_{0}^{2}}=\frac{(1+r_{0})^{2}}{n_{0}r_{0}^{2}}
 \end{equation}
where the addition of the term $r_{p}$ to the radius $R_{0}$ of the grain in the second term of the equation stems from the fact that any particle which center is located outside the grain but at a distance lower than $r_{p}$ intersects with the surface of the grain. Now the probability $p(X^{i}=k)$ for the grain $i$ to contain exactly $k$ particles is: 

\begin{equation}
 p(X^{i}=k)={N_{p}\choose k}\left(\frac{(1+r_{0})^{2}}{n_{0}r_{0}^{2}}\right)^{k}\cdot \left( 1-\frac{(1+r_{0})^{2}}{n_{0}r_{0}^{2}}\right)^{N_{p}-k}
 \end{equation} 

Eventually, assuming that every contracting grain that initially intersects less than three particles disappears, the probability for a contracting grain to vanish can be calculated as: 

\begin{equation}
p_{v}= \lim_{n_{0}\to\infty}  \sum_{k=0}^{2}{N_{p}\choose k}\left(\frac{(1+r_{0})^{2}}{n_{0}r_{0}^{2}}\right)^{k}\cdot \left( 1-\frac{(1+r_{0})^{2}}{n_{0}r_{0}^{2}}\right)^{N_{p}-k} 
\end{equation}  
 where the total number of particles in the microstructure can be calculated as a function of $n_{0}$ as $N_{p}= n_{0}f_{s}r_{0}^{2}$. 

\subsection{Mean grain size in a pinned microstructure}
Knowing the equilibrium size of the grains that have grown and of those that have shrunk, as well as the fraction of grains from the initial microstructure that grow, shrink and vanish, respectively, the mean grain size in the stabilised microstructure can be calculated. The fractions $x^{lim}
_{g}$ and $x^{lim}
_{s}$ of large and small grains in the final microstructure are given by: 
\begin{subequations}
\begin{empheq}{align}   
x^{lim}_{g}=\frac{x^{0}_{g}}{x^{0}_{g}+x^{0}_{s}} \\
x^{lim}_{s}=\frac{x^{0}_{s}}{x^{0}_{g}+x^{0}_{s}}
\end{empheq}
\end{subequations}
and finally the mean normalised limiting size for grains in the pinned microstructure is :
\begin{equation}
\bar{r}_{lim}=\sqrt{x^{lim}_{g}r_{g}^{2}+ (1-x^{lim}_{g})r_{s}^{2}}
\end{equation}
\section{Computer simulations}

\subsection{Computer model}
Grain growth in 2D polycrystals containing particles is simulated thanks to a finite element level set formulation that has been detailed elsewhere \cite{Osher1988,Osher2001, Osher2004, Agnoli2014}. This method represents interfaces between grains via distance functions, defining contours or surfaces within the simulation domain, where the isovalue zero corresponds to the interface. The evolution of each grain's interface follows a kinetics equation defined as:
\begin{equation}
\vec{v}=M P \vec{n}
\end{equation}
where $\vec{n}$ is the outside unitary normal and $P=-\gamma\kappa$, with $\kappa$ the local curvature.

Particles are introduced into the finite element mesh as voids, with their interaction with grain boundaries considered via imposed boundary conditions, specifically the contact angle between the boundary and the particle. An advantage of this formulation is its explicit consideration of particle-grain boundary interaction, eliminating the need for computing an equivalent pressure exerted by particles on grain boundaries. Consequently, each interface naturally evolves towards locally reducing its surface (or length in two dimensions). 
 
\subsection{Microstructure generation}
The initial microstructures are generated using Laguerre Voronoï tesselation with an existing tool described in \cite{Hitti2012}, so as to build polycrystals with controlled average initial size $\bar{R_{0}}$.  
\subsection{Simulation set-up}
The simulations were conducted across a range of particle surface fractions $f_{s}$, spanning from 1\% to 40\%, and initial normalized grain sizes $r_{0}$, spanning from 1 to 40. Details regarding the specific values of $f_{s}$ and $r_{0}$ used in the simulations can be found in \ref{appendix:simulations}. Each simulation proceeded until a stable configuration was attained, meaning when the microstructure no longer evolves. The domain size was selected to retain a statistically representative number of grains at the end of each simulation. The numbers of grains in the pinned configurations are reported in \ref{appendix:simulations}. 

The level set formulation used in this work \cite{Agnoli2014} requires as an input the grain boundary mobility as well as the grain boundary interfacial energy. The interfacial energy is considered here as isotropic, and since the present work focuses on the equilibrium configuration, which is neither affected by the value of the interfacial energy nor by that of the grain boundary mobility, these two parameters have been arbitrarily set. 

At the end of the simulation, the average normalised grain size of the grains that have grown is calculated as: 

\begin{equation}
\bar{r}_{g}=\sqrt{\frac{\sum\limits_{S_{i}>S_{0}}{S_{i}}}{\pi n_{g}}}
\end{equation}
and the average normalised grain size of the grains that have shrunk is calculated as:
\begin{equation}
\bar{r}_{s}=\sqrt{\frac{\sum\limits_{S_{i}\leq S_{0}}{S_{i}}}{\pi n_{s}}}
\end{equation}

\section{Results and discussion}

\subsection{Simulation results}
\begin{figure}[H]

\begin{subfigure}{0.5\textwidth}
\includegraphics[trim={0cm 0cm 0cm 3cm},clip, scale=0.3]{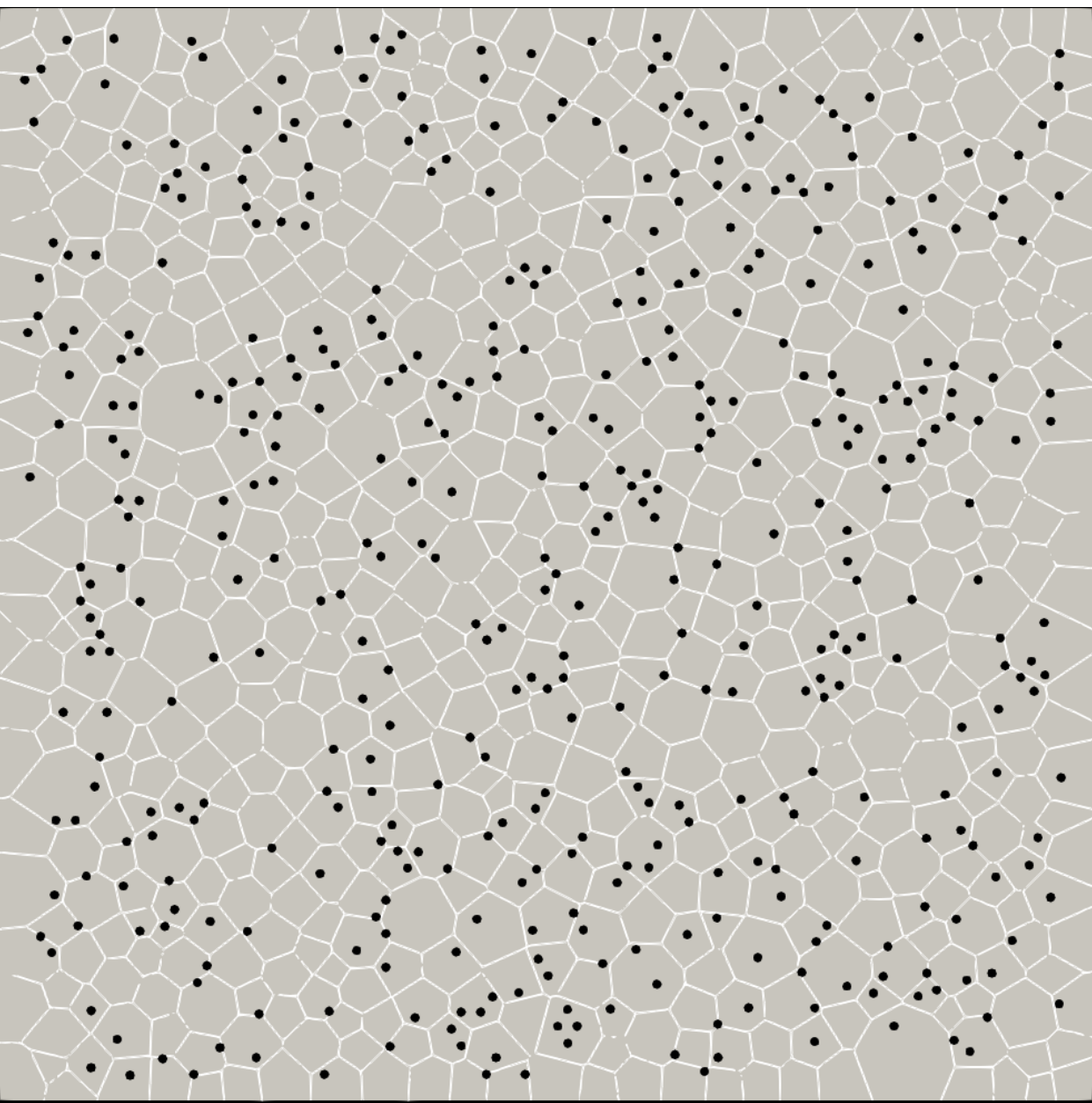} 
\caption{}
\label{subfig:initial_microstructure}
\end{subfigure}
\hfill
\begin{subfigure}{0.5\textwidth}
\includegraphics[trim={0cm 0cm 0cm 3cm},clip, scale=0.3]{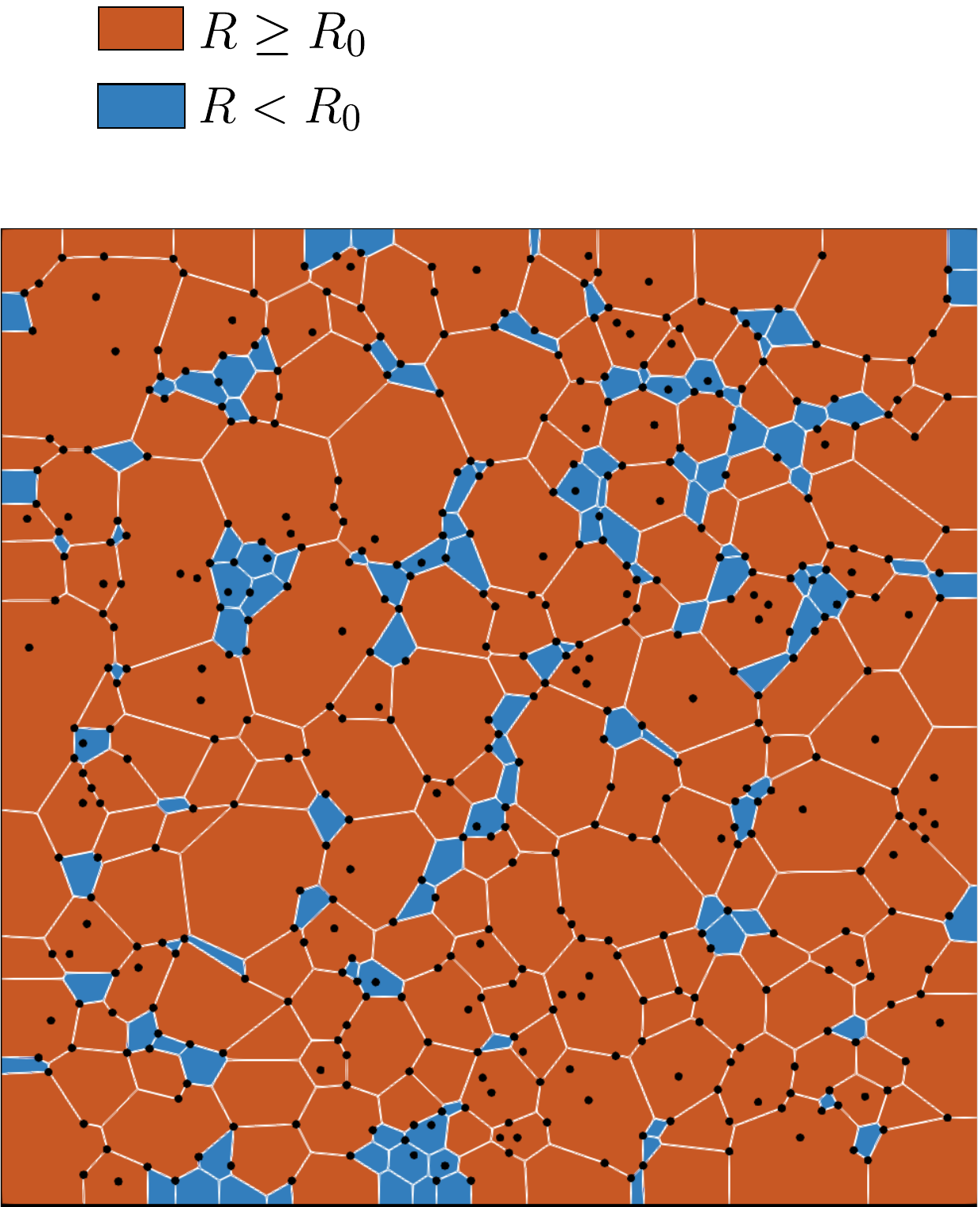}
\caption{}
\label{subfig:stabilised_microstructure}
\end{subfigure}
\caption{Microstructure evolution for $f_{s}$=3\% and $r_{0}$=4, showing (a) initial microstructure; (b) stabilised microstructure. The white lines denote the grain boundaries, the black circles represent the particles, and the grain colours on (b) denote the grains that have grown or shrunk.}
\hfill
\label{fig:microstructures}
\end{figure}

In \fref{fig:microstructures}, an example of a simulated microstructure evolution is shown, with the initial microstructure displayed in \fref{subfig:initial_microstructure}, and the stabilised microstructure in \fref{subfig:stabilised_microstructure}. In \fref{subfig:stabilised_microstructure}, the different colours indicate the grains that have grown and those that have shrunk with respect to the initial mean grain size.

\subsection{Model validation}\label{subsection:model_validation}

\begin{figure}[H]
\begin{subfigure}{0.55\textwidth}
\includegraphics[trim={2cm 0cm 2cm 0cm},clip, scale=0.65]{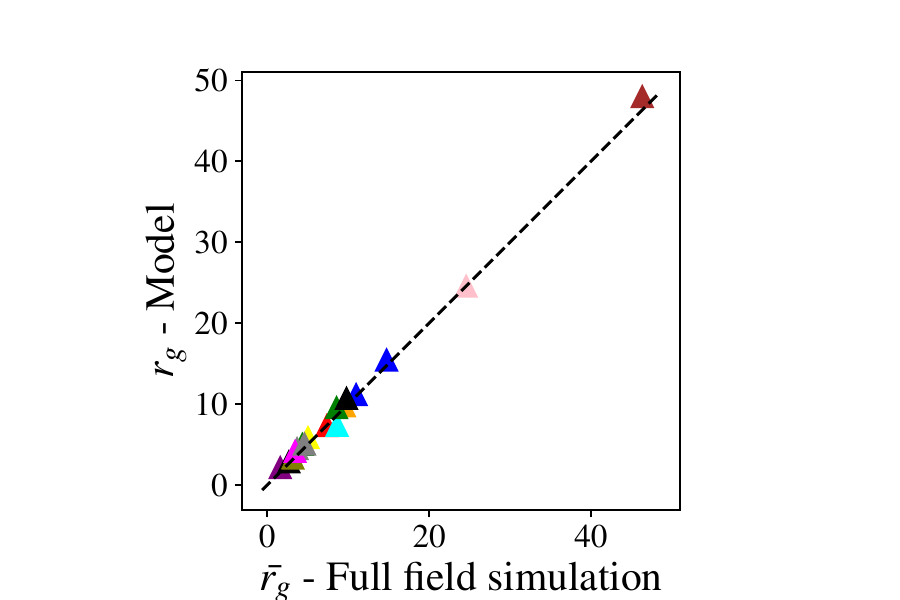} 
\caption{}
\label{subfig:results_growing_grain}
\end{subfigure}
\hfill
\begin{subfigure}{0.55\textwidth}
\includegraphics[trim={2cm 0cm 2cm 0cm},clip, scale=0.65]{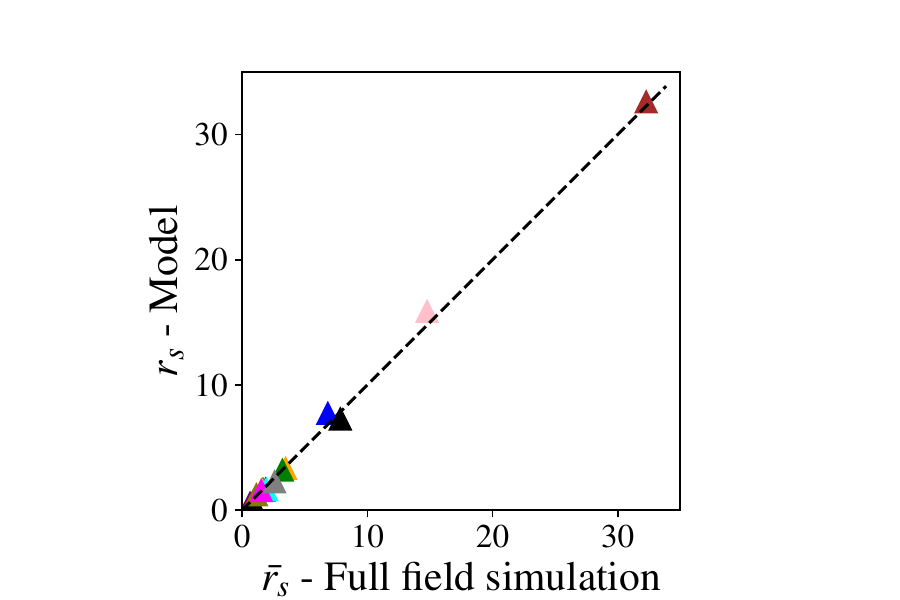} 
\caption{}
\label{subfig:results_shrinking_grain}
\end{subfigure}
\hfill
\center
\begin{subfigure}{0.55\textwidth}
\includegraphics[trim={2cm 0cm 2cm 0cm},clip, scale=0.65]{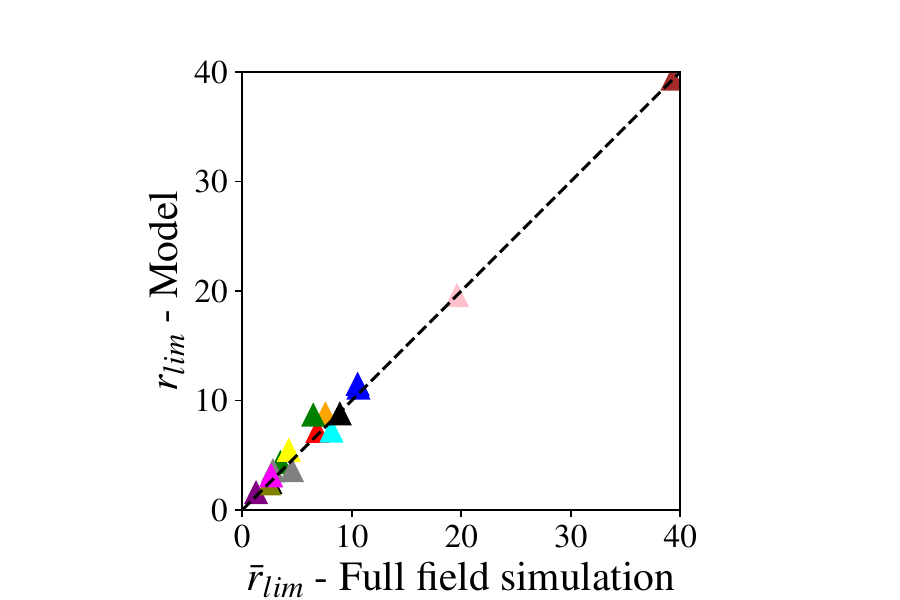} 
\caption{}
\label{subfig:results_shrinking_grain2}
\end{subfigure}
\caption{Limiting grain size, calculated with the model presented here, as a function of limiting grain size obtained in the full-field simulations; (a) calculated values for limiting grain size of growing grains; (b) calculated values for limiting grain size of shrinking grains; (c) calculated values for mean limiting grain size. The simulations cover various ranges of $r_0$ and $f_{s}$ and each point  in the chart is the result of a single simulation. The dashed line shows the "$y=x$" line.  }
\label{fig:results_grain_size}
\end{figure}

In order to validate the model described  \sref{section:analytic_model}, \fref{fig:results_grain_size} displays the calculated values for the normalised grain size, compared to the results derived from full-field simulations. Each data point in the chart corresponds to one of the conditions outlined in \ref{appendix:simulations}. The comparison depicted in \fref{fig:results_grain_size} demonstrates very good agreement between the model predictions and the full-field simulation results. This holds for the stagnated normalised sizes of growing and shrinking grains, as well as for the averaged normalised grain size in the stabilised microstructure.

It can be concluded that the simple analytical model we propose here, which is free of any fitting parameter, can reliably predict the results derived from computationally costly full-field simulations. Additionally, it is noteworthy that while most models in the literature rely on the assumption that the grain size is significantly larger than the particle size \cite{Smith1948, Hellman1975, Worner1987, Louat1982}, our model does not require such a restriction, and remains valid even when grains and particles have similar sizes. More detailed comparison with previous models from the literature will be discussed in \sref{subsection:comparison_with_literature}.

\subsection{Hypotheses on grain shape}
\begin{figure}[H]
\begin{subfigure}{0.55\textwidth}
\includegraphics[trim={2cm 8cm 0cm 9.5cm},clip, scale=0.45]{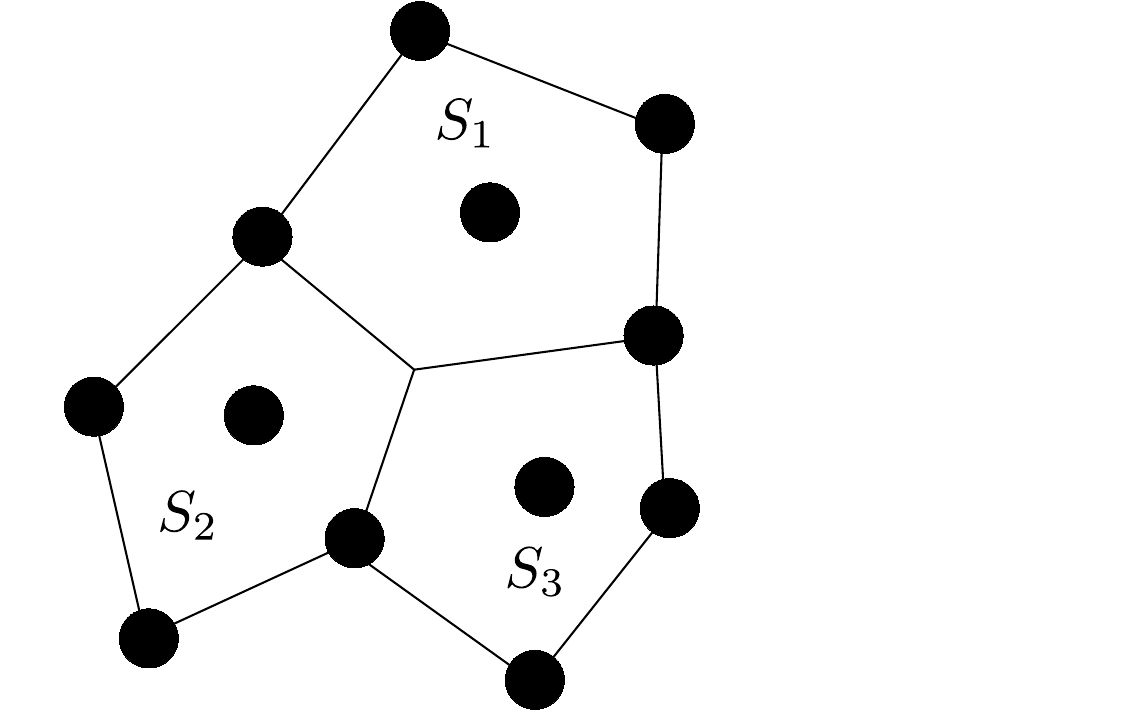} 
\caption{}
\label{subfig:triple_point_real}
\end{subfigure}
\hfill
\begin{subfigure}{0.55\textwidth}
\includegraphics[trim={2cm 8cm 0cm 9.5cm},clip, scale=0.45]{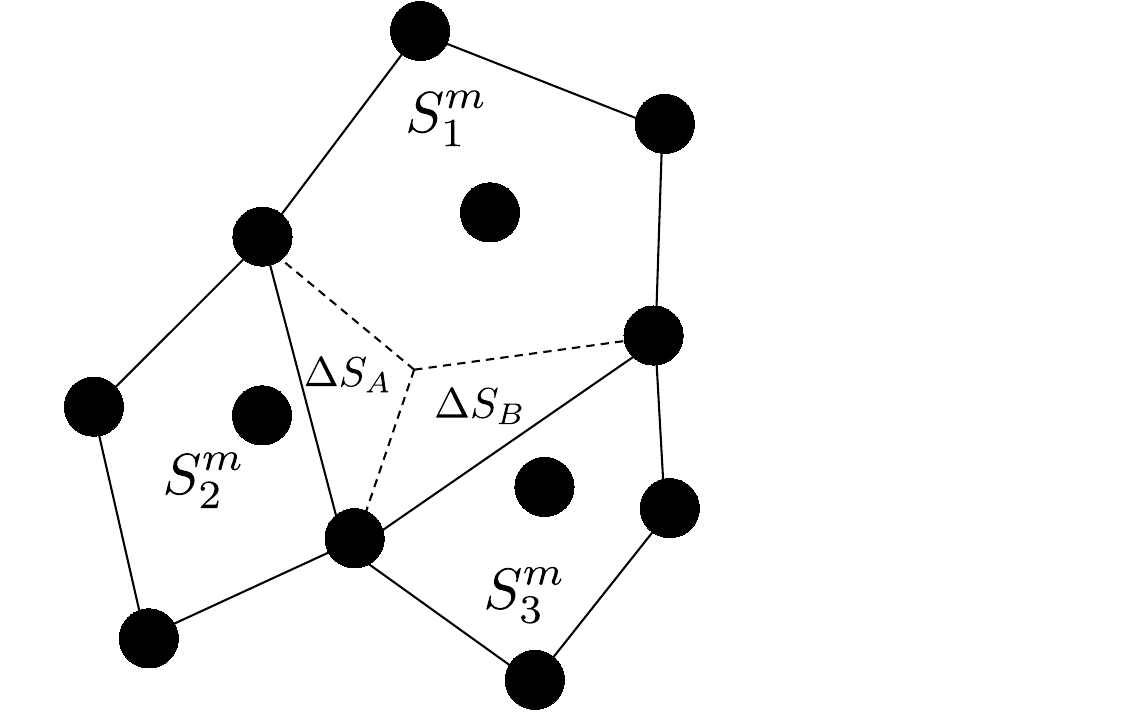} 
\caption{}
\label{subfig:triple_point_model}
\end{subfigure}

\caption{(a) configuration of a grain with particles and a triple junction (b) configuration of a grain ignoring triple junctions. The plain lines indicate the grain boundaries. }
\label{fig:schematic_triple_points}
\end{figure}

The analytical model presented here relies on several simplifying assumptions, notably regarding the grain shape. Most importantly, the motion of triple junctions has been ignored in the present analysis. It is somewhat surprising that the model reproduces so well the simulation data (\sref{subsection:model_validation}) despite ignoring triple junctions. A possible interpretation for this is shown in \fref{fig:schematic_triple_points}, where \fref{subfig:triple_point_real} shows a three-grains configuration with a triple junction at equilibrium (120°/120°/120° angles) and \fref{subfig:triple_point_model} shows the consequence of ignoring this triple junction equilibrium in the model. Let $S_{1}$, $S_{2}$ and $S_{3}$ be the respective surfaces of the three grains displayed in \fref{subfig:triple_point_real}; and $S_{1}^{m}$, $S_{2}^{m}$ and $S_{3}^{m}$ the surfaces of the grains in the second configuration (\fref{subfig:triple_point_model}), where the triple junction is ignored. It is clear from the figure that $S_{1}^{m}=S_{1}+\Delta S_{A}+\Delta S_{B}$, $S_{2}^{m}=S_{2}-\Delta S_{A}$, and 
$S_{3}^{m}=S_{2}-\Delta S_{B}$. As a result, calling $\bar{S}$ the mean surface of the grains in the first configuration, and $\bar{S_{m}}$ that in the second, it is clear that $\bar{S}=\bar{S_{m}}$. This could help explain why ignoring the triple junction does not result in significant error regarding limiting radius calculation. Neglecting the triple junctions might also affect the calculation of the number of vanishing grains. Indeed, while, in our model, the grains that intersect two particles are expected to vanish, some of the grains intercepting two particles might in reality also display an equilibrium triple junction and thus get stabilised. This point will be discussed just below.

Another key assumption of the model, somewhat related to the fact that triple junctions are ignored, is that the equilibrium shape of a grain - once the stable microstructure configuration has been reached - is considered as a regular polyhedron with one particle at each vertex, and with side length equal to the average free path between particles.

In order to verify this hypothesis, the stable microstructure reached at the end of the simulation was analysed in more details for six arbitrarily chosen simulations. For each of these simulations, all grains $i$ present in the microstructure were listed, along with their normalised surface $s_{i}$, defined as $s_{i}=S_{i}/(\pi r_{p}^2)$ and the number of particles meeting their interface $N_{i}$. This information was then averaged on integer values of $N$, allowing to compute, for a given number of precipitates $N$, the average normalised surface of the grains having $N$ precipitates meeting their boundary. 

\begin{figure}[H]
\centering
\includegraphics[trim={0cm 0cm 0cm 0cm},clip, scale=0.6]{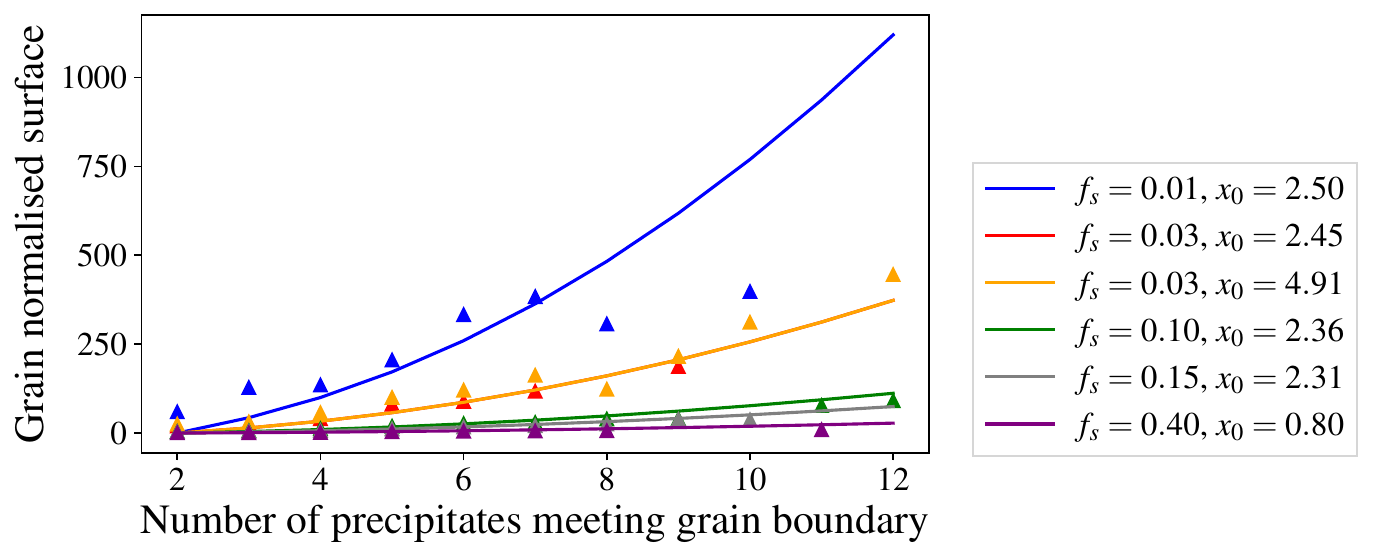} 
\caption{Average normalised grain surface as a function of the number of particles meeting the grain boundary. The plain lines indicate the results obtained with the model (each line corresponds to a set of $r_{0}$ and $f_{s}$) and the markers indicate the values, obtained for individual grains in the full field simulations, grouped by values of $N$ and then averaged. }
\label{fig:N_grain_area}
\end{figure}

The grain normalised surface, calculated using the analytical model as a function of the number of precipitates meeting its boundary, is shown in \fref{fig:N_grain_area}, along with the results of the full-field simulations. As can be seen in the figure, decent agreement is obtained between the analytical model and the full-field simulations. It can also be observed that, as previously mentioned, some grains displaying only two particles at the boundary might still be present in the pinned microstructure, although our model considers that such grains should not be stabilised. However, this has little influence on the mean grain size calculation, as demonstrated by the good agreement between simulations and analytical model. Additionally, it will be shown later that the trends exhibited by the number of growing, shrinking, and vanishing grains, can be be reasonably predicted. confirming that the hypotheses made in our analytical model, which allow considerable simplification of the problem, are reasonable.  This suggests that the hypotheses made on grain shape, which allow considerable simplification, are reasonable. 

\subsection{Effect of initial grain size on subsequent growth}
\begin{figure}[H]

\center
\includegraphics[trim={0cm 0cm 0cm 0cm},clip, scale=0.65]{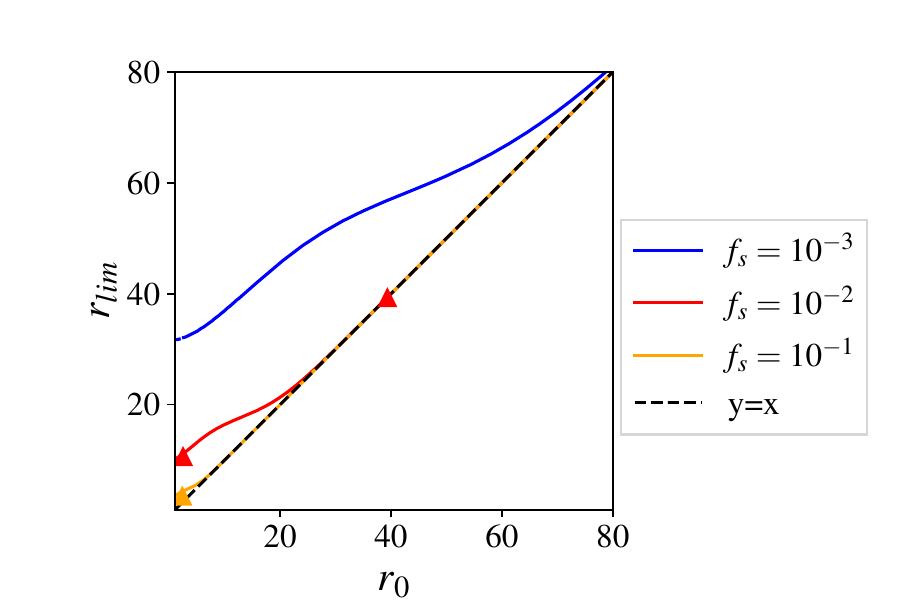} 
\caption{Calculated mean normalised size in the equilibrium configuration with respect to the initial normalised size. The plain lines indicate the results obtained with the model while the individual points show some of the results of the full-field simulations. The dashed line shows the "$y=x$" line. }
\label{fig:initial_size}
\end{figure}

An important feature of the analytical model presented in this work is that it takes into account the effect of the initial grain size on the subsequent microstructure development. This effect is depicted in \fref{fig:initial_size} where the influence of the initial grain size on the limited grain size is illustrated. The figure shows that, for a fixed particle surface fraction, the normalised grain size obtained in the stable microstructure increases when the initial normalised grain size increases. While the relationship between the initial normalised grain size and $r_{lim}$ exhibits near-linearity at high particle surface fractions (1\% and 10\%), non-linear behaviour is seen at lower surface fractions (0.1\%), especially for small initial normalised grain size. More specifically, the initial normalised grain size has less influence on the stagnated grain size if the particle surface fraction is low and if the initial normalised grain size is small. This aligns with some of the results of a  phase-field study by Moelans and co-authors~\cite{Moelans2006}, who observed that the role of the initial grain size is increasingly important for large initial grain size and low particle fractions. 

Most analytical models in the literature, such as the Smith-Zener model and its various extensions or modifications, have relied solely on the particle surface fraction as an input for predicting stagnated normalised grain size \cite{Haroun1968, Louat1982, Nes1985, Hillert1988, Hunderi1989, Hazzledine1990, Worner1993}. However, our present work reveals that this input alone seems insufficient for accurate prediction. This observation is in line with findings in the literature, since it has for example been reported that different initial microstructures with identical particle surface fractions exhibit varying stagnated normalised grain sizes \cite{Nishizawa1997, Agnoli2014, Moelans2006}.  Our current study demonstrates the importance of considering the initial normalised grain size as another essential input for predicting limiting grain size. Previous research has also used simulation tools to point an influence of initial grain size on limiting grain size \cite{Weygand1999, Hunderi1982, Moelans2006, Payton2013}. Nonetheless, to the best of our knowledge, a concise expression for calculating stagnated normalised grain size as a function of initial grain size has not been proposed before. Therefore, our work not only underscores the critical role of accounting for initial grain size, but also introduces a straightforward method for estimating limiting normalised grain sizes based on the initial normalised grain size.

\subsection{Effect of particles on stagnated grain sizes}

\begin{figure}[H]

\center
\includegraphics[trim={0cm 0cm 0cm 16cm},clip, scale=0.45]{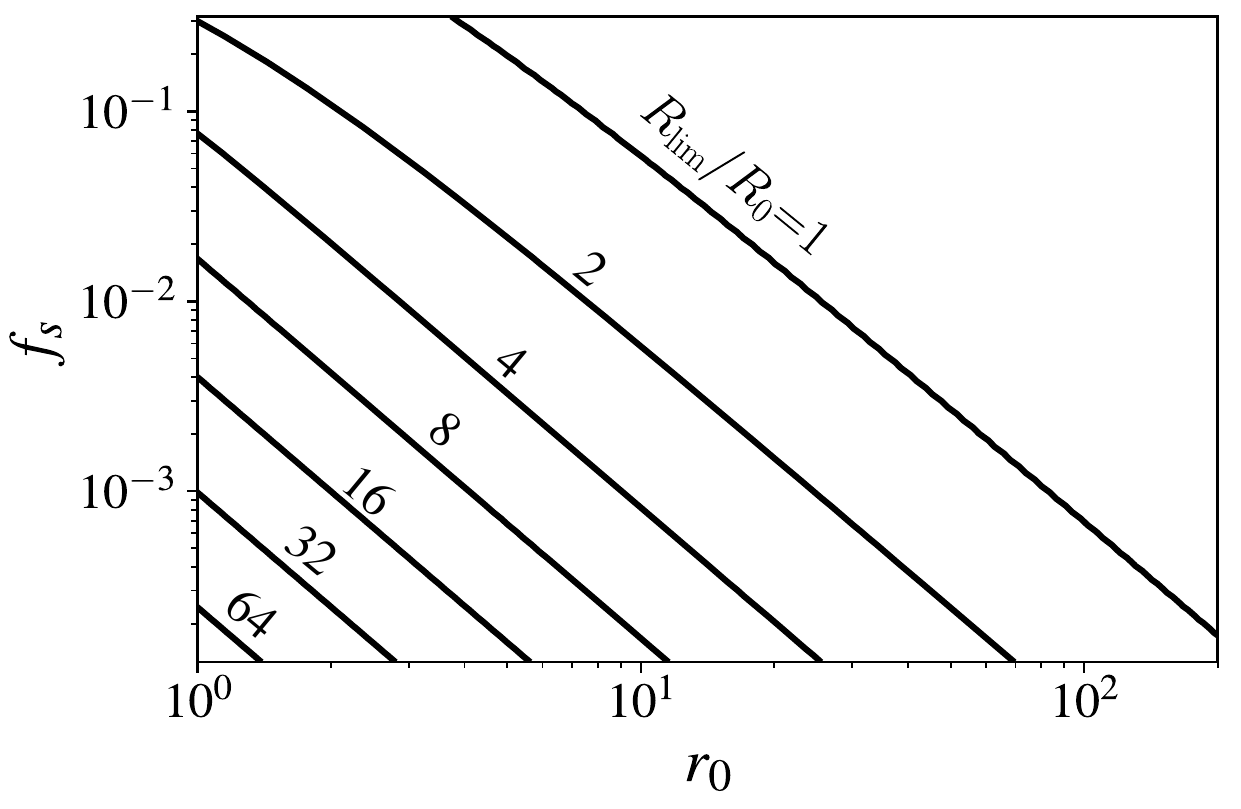} 
\caption{Isolines for the estimated limiting grain size divided by the initial grain size as a function of particle surface fraction and initial normalised grain size. The number along the lines indicate the iso-values for $R_{\rm lim}/R_{0}$.}
\label{fig:isolines}
\end{figure}

The present analytical model offers a simple way to calculate the limiting grain size $R_{\rm \lim}$ in pinned microstructures as a function of the initial grain size and of the particle surface fraction. Isolines illustrating the expected ratio between stagnated grain size and initial grain size are depicted in \fref{fig:isolines}, as a function of both the particle surface fraction and the initial normalised grain size. In \fref{fig:isolines}, the efficiency of grain boundary pinning by second-phase particles is observed to be most pronounced at elevated particle surface fractions and large initial normalised grain sizes. This phenomenon is evidenced by the stagnated grain size converging towards the initial grain size ($R_{lim}/R_{0} \simeq 1$), which indicates minimal microstructural evolution.

Furthermore, maintaining a constant absolute initial grain size ($R_{0}$) and particle volume fraction, the model indicates that the pinning effect is enhanced with decreasing particle size (since $r_0=R_{0}/r_{p}$)

These observations align with established literature, where the ratio $f_{s}/r_{p}$ commonly serves as an indicator of pinning effect strength \cite{Manohar1998}. Conversely, the isolines in \fref{fig:isolines} demonstrate that under conditions of low initial normalised grain size and low particle surface fraction, maximal growth is exhibited.

\subsection{Effect of particles on grain size heterogeneity} \label{subsection:heterogeneity}

The model proposed here relies on the observation that particles not only impinge grain growth but also prevent - to some extent - shrinking grains from vanishing. This results in a bimodal grain distribution made of a mixture of small and large grains, whose respective sizes and fractions can be calculated. The model can thus be used to estimate the grain size heterogeneity induced by the presence of particles. 

\begin{figure}[H]
\centering
\includegraphics[trim={0cm 0cm 0cm 0cm},clip, scale=0.7]{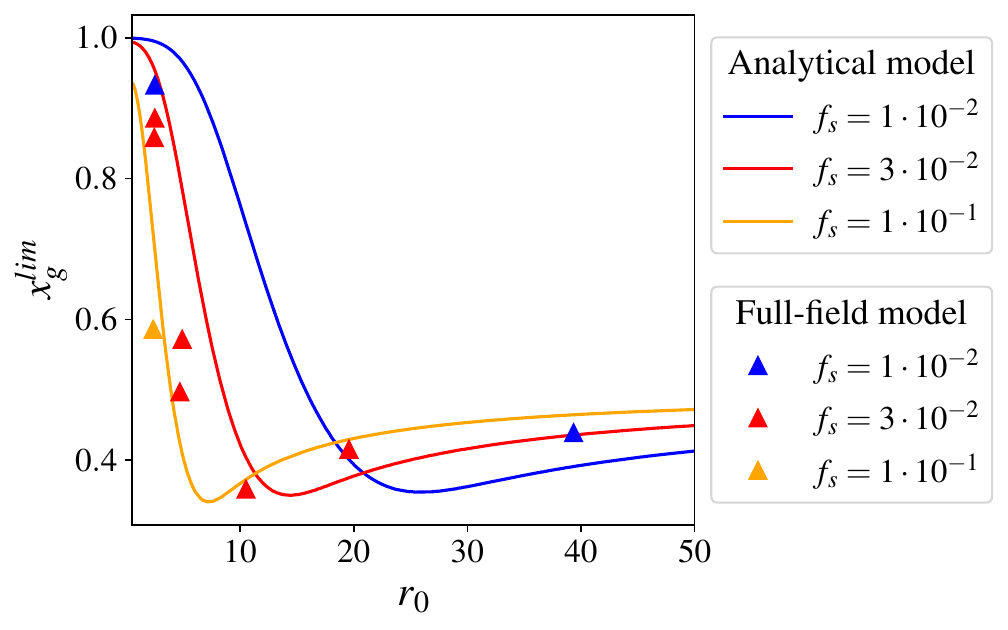} 
\caption{Calculated fraction of large grains in the pinned microstructure. Both results derived from the analytical model and from full-field simulations are displayed.   }
\label{fig:r0_xlimg}
\end{figure}

The influence of the initial normalised grain size on the expected fraction of large grains in the stabilised microstructure, along with results derived from full-field simulations, is shown in \fref{fig:r0_xlimg}. It can first be seen in the figure that our analytical model reasonably reproduces the trends observed in full-field simulation.

\Fref{fig:r0_xlimg} shows that, when the initial normalised grain size is low, the fraction of large grains is close to 1, meaning that there are only large grains in the stabilised microstructure. The reason for this is that, when the initial grain size is small, the probability that a grain surface initially intercepts three particles or more is very low. Thus, all the grains that contract are able to disappear, and the only remaining grains in the microstructure are those that have grown. Conversely, when the initial normalised grain size becomes high, it becomes increasingly likely that the surface of a grain initially intercepts three particles. Thus, in this situation, no grain vanishes, and, as a result, no grain can significantly grow. In this situation, only limited grain boundary motion is possible, and, considering that in the initial state, grains are equally likely to grow or shrink, $x_{g}^{lim}$ converges towards 1/2. One should keep in mind, however, that when the initial normalised grain size is high, the notion of "large" and "small" grains becomes less relevant, since grown and contracted grains have similar sizes. This aspect will be detailed in the following. 

\begin{figure}[H]
\centering
\includegraphics[trim={0cm 0cm 0cm 0cm},clip, scale=0.7]{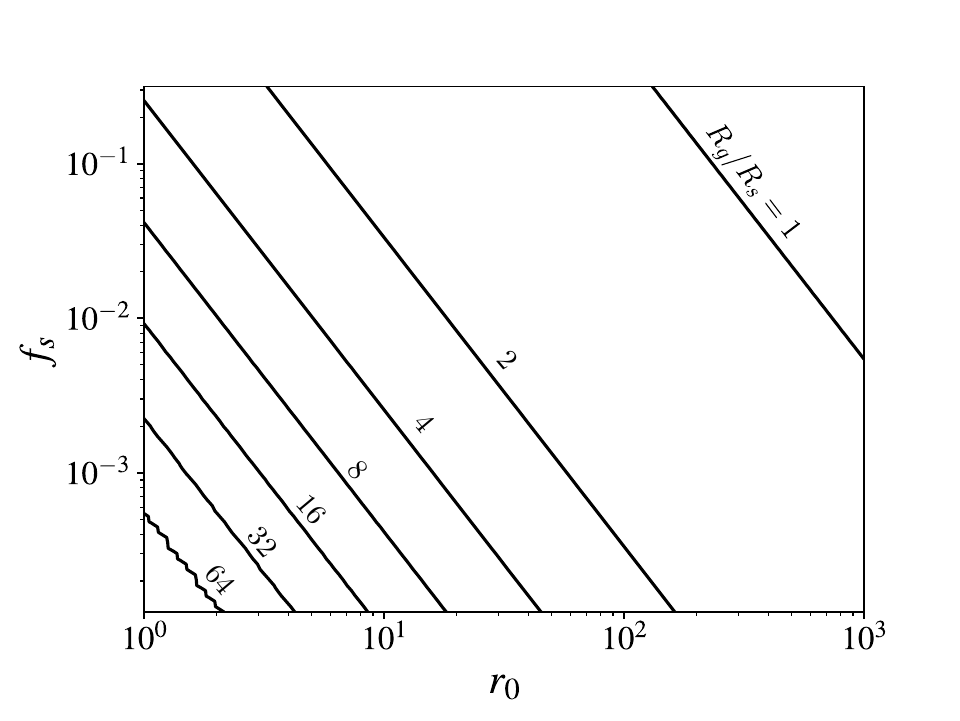} 
\caption{ Isolines for the calculated size ratio between large and small grains in the stabilised microstructures, with respect to initial normalised grain size and particle surface fraction. The number along the lines indicate the iso-values for $R_{g}/R_{s}$. }
\label{fig:isoline_heterogeneity}
\end{figure}

In order to predict microstructure heterogeneity, it is useful to compare the respective sizes of large and small grains within stabilised microstructures. The isolines in~\fref{fig:isoline_heterogeneity}, depict the calculated ratio $R_{g}/R_{s}$ between the sizes of large and small grains. The figure reveals that lower particle surface fractions and smaller initial normalised grain sizes induce higher size differences between so-called large and small grains. 

\begin{figure}[H]
\centering
\includegraphics[trim={0cm 0cm 0cm 0cm},clip, scale=0.8]{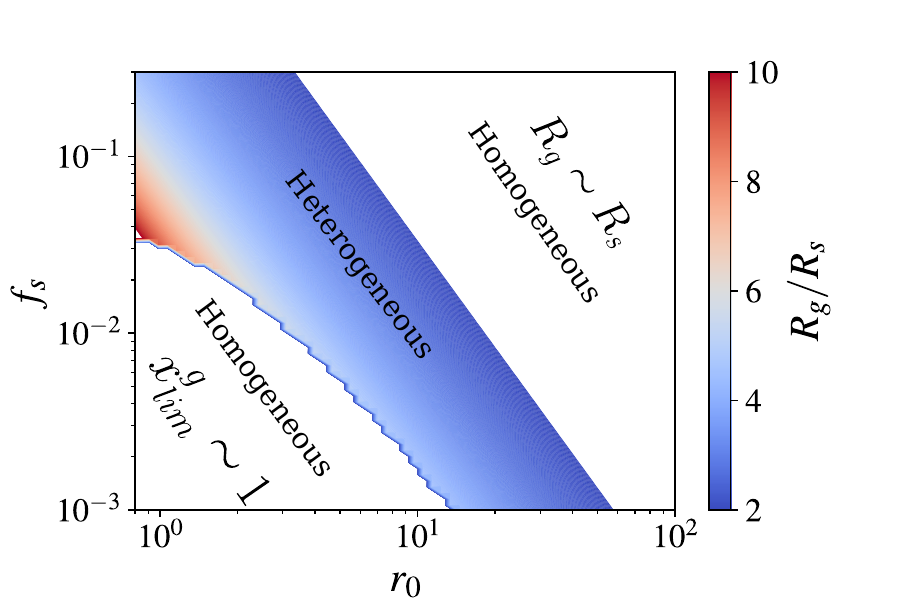} 
\caption{Regimes in the ($r_{0}$, $f_{s}$) space where the grain size is expected to be heterogeneous or homogeneous. The coloured areas corresponds to values where $R_{g}$/$R_{s}>2$, and where the fraction of small grains in the microstructure is higher than 1\%. The colour gradient corresponds to varying values of $R_{g}$/$R_{s}$ }
\label{fig:heterogeneity}
\end{figure}

To predict the apparition of grain size heterogeneity resulting from particle pinning, the calculated values of the grain sizes and fractions of large and small grains should be considered concomitantly. Indeed, the analysis above shows that the values of $r_{0}$ and $f_{s}$ leading to the highest size difference between small and large grains are also regimes in which the fraction of small grains in the stabilised microstructure is very low. A criterion to distinguish regimes where grain size heterogeneities might develop is proposed in \fref{fig:heterogeneity}. In that figure, a microstructure is considered as heterogeneous if $R_{g}$/$R_{s}>2$, and if the fraction of small grains is larger than 1\%. The figure shows that two different regimes might lead to homogeneous microstructure. If the particle surface fraction and initial normalised grain size are high, limited evolution in the grain size is observed (as previously discussed), and the grain size distribution remains fairly homogeneous. Conversely, when the particle surface fraction and initial normalised grain size are low, it is unlikely that contracting grains get pinned by particles, which results in a homogeneous microstructure made of large grown grains. In-between these two regimes, \fref{fig:heterogeneity} shows that some heterogeneity might develop in the grain size distribution, with large stabilised grains up to 10 times larger than small stabilised grains.

The development of grain size heterogeneity due to the presence of particles may have important practical consequences. It is indeed well-known that grain size plays a crucial role in in-service properties of metallic materials. Polycrystals with uniform grain size are often desirable \cite{Cottrell1949}, and in non-uniform microstructures, the performances of the material are usually limited by its weakest points. While large grains reduce strength as well as fracture toughness \cite{Hull2011, Cottrell1949}, small grains may be detrimental to ductility and creep resistance \cite{kassner2004fundamentals}. Thus, grain size is usually targeted according to the desired application, and mixtures of small and large grains are usually to be avoided. Although comparison between 3D real situations and 2D models should be treated cautiously, the present findings could indicate that particles, in some specific regimes of initial grain size and particle volume fraction (\fref{fig:heterogeneity}),  may produce microstructures with non-uniform grain size, and thus might be responsible for degraded mechanical properties. 

The present work points out that, in addition to limiting the growth of large grains, particles can also prevent small grains from disappearing, thus resulting in bimodal grain size distributions. The presence of such small grains in pinned microstructures has also been stressed by Srolovitz \cite{Srolovitz1984}, who performed full-field Monte Carlo simulations of grain growth in the presence of particles. Nonetheless, to the best of our knowledge, no analytical model from the literature considered this effect before.  The capability to predict the respective fractions and sizes of small and large grains due to particle pinning may help to predict and thus control microstructure development. 

\subsection{Comparison with existing models}\label{subsection:comparison_with_literature}

As already stated in the introduction of this work, the investigation of particle pinning has been the focus of considerable literature. While earlier approaches have demonstrated success in capturing general trends in limiting grain size, no analytical model to date allows to predict accurately particle pinning.  A common feature among most analytical models from the literature is the assumption that the boundary is able to break away from precipitates when $ P_{\gamma} >P_{z}$, with   $P_{\gamma}\propto \gamma/R_{g} $. 

The motivation behind this work stems from the observation that, in the absence of a free energy difference between adjacent grains, grain boundaries should not spontaneously break free from precipitates, as detailed in \sref{section:analytic_model}. The calculation of the driving pressure for grain boundary motion using an expression of the form $P_{\gamma} \propto \gamma/R_{g}$ might actually be questionable as soon as the boundary is separated in several segments by particles. In a two-dimensional microstructure, two segments of the same grain boundary separated by particles are no longer related to each other, and are thus unlikely to move in a coordinated way. The fact that grain boundaries cannot by-pass particles without external driving pressure (such as deformation energy) is implicit in the treatment of Ashby and Lewis \cite{Ashby1968}, and has been pointed out by Kim and Kishi \cite{Kim1999}. Gladman calculated the energy barrier for particle unpinning and concluded that particle by-passing was unlikely to occur by thermal activation \cite{Gladman1966}. However, existing analytical models do not account for this aspect when computing the limiting grain size in pinned microstructures.

\begin{figure}[H]
\centering
\includegraphics[trim={0cm 0cm 0cm 0cm},clip, scale=0.55]{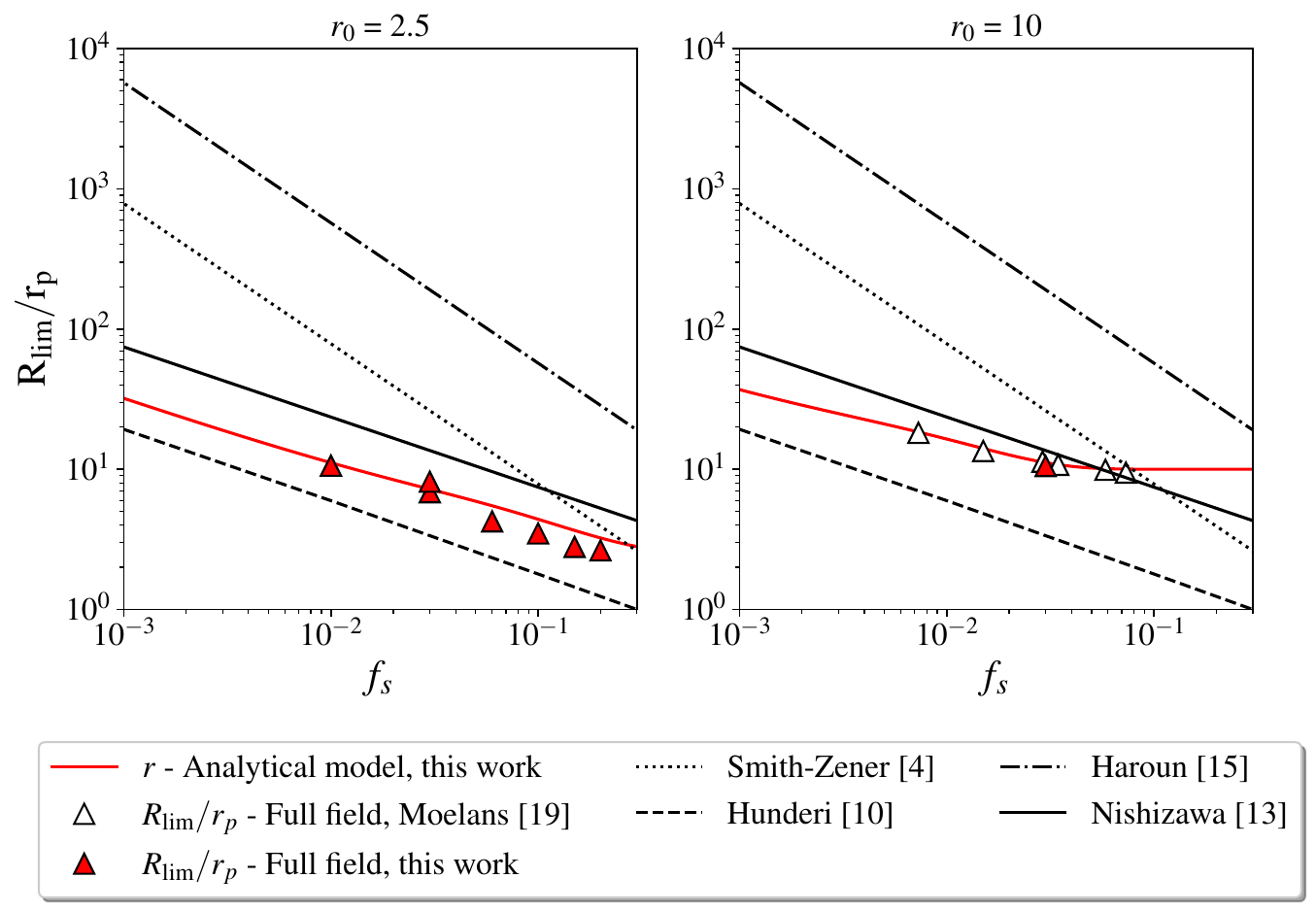} 
\caption{Comparison between the limited grain sizes calculated with the analytical model of the present work and the results obtained with models from literature. The results from full-field simulations of this study are also displayed.}
\label{fig:comparison_models}
\end{figure}

A comparison between the results obtained with the present analytical model and those obtained with existing models from the literature\footnote{Three of the four evaluated models from the literature were originally developed in three-dimensions and were thus converted in two-dimensions to allow comparison with our present model} is shown in \fref{fig:comparison_models}, along with results from full-field simulations conducted in this study. Results from the phase field simulations by Moelans \textit{et al.} \cite{Moelans2006} are also displayed for comparison. 

Comparison of our model with existing literature models is illustrated in \fref{fig:comparison_models}, for two arbitrarily chosen initial normalised grain sizes. Our model yields distinct results compared to those from the literature, demonstrating better agreement with results derived from full-field simulations conducted in this study. Notably, none of the analysed models from the literature can explain the results derived from the full-field simulations of the present work. Furthermore, \fref{fig:comparison_models} highlights the consistency of our analytical model not only with the simulations conducted in this study but also with simulations reported in the literature. Particularly, in the stagnated microstructure, the mean normalized grain size obtained by Moelans and co-authors \cite{Moelans2006} using phase-field simulations is very-well predicted by our analytical model, as depicted in \fref{fig:comparison_models}.

All models from the literature analysed in this work predict a linear or almost-linear dependency between the logarithm of $R_{\rm{lim}}/r_{p}$ and that of $f_{s}$, in line with the well-accepted relation $R_{\rm{lim}}/r_{p}=\beta/f_{s}^{m}$, with $\beta$ and $m$ being constants. However, our analytical model does not exhibit such linearity. Given that our analytical model better reproduces the results of full-field simulations compared to existing models from the literature, this suggests that the relation $R_{\rm{lim}}/r_{p}=\beta/f_{s}^{m}$ might not be the most appropriate to describe particle pinning in two-dimensional polycrystals.

\section{Conclusions}

In this study, we developed a parameter-free analytical model to predict stagnated grain size in two-dimensional microstructures anchored by second-phase particles, but also the resulting grain size heterogeneity. The model capitalises on the observation that, in the absence of free energy variation between adjacent grains, a grain boundary segment confined by two particles assumes a straight-line configuration and cannot bypass the particles. By assuming grains form regular polygons, the restricted grain size can be readily calculated by enumerating the number of particles intercepted by the grains during their growth or contraction. 

Validation of the analytical model was achieved through level-set full-field simulations, which corroborated findings from existing literature. Notably, the model demonstrates strong agreement for particle surface fractions ranging from 1\% to 40\% and across a wide range of initial grain sizes, even when the initial grain size is comparable to the particle size.  

The simple analytical model presented here allows to rationalise some features of particle pinning, such as the influence of  initial grain size, or the development of heterogeneous grain size distribution, that had, to the best or our knowledge, only been investigated through computationally heavy full-field simulations so far. The key conclusions derived from this study are as follows:

\begin{itemize}
   \item{The initial normalised grain size significantly influences the prediction of stagnated grain size, and the model introduced herein offers insights into its influence.}
   \item{ Particles not only inhibit grain growth but also impede grain contraction, possibly leading to a heterogeneous distribution of grain sizes within microstructures, characterised by a mixture of small and large grains. While, when particle surface fraction and initial normalised grain size are either very low or very large, the grain distribution is homogeneous, a regime of intermediate values of $f_{s}$ and $r_{0}$ has been identified where this heterogeneity becomes significant. }
   \item{Although particle pinning becomes increasingly effective for smaller particles and higher particle surface fractions, the relationship $R_{\rm{lim}}/r_{p}=\beta/f_{s}^{m}$ fails to adequately describe stagnated grain size in two-dimensional microstructures.}

\end{itemize}

Future research will focus on adapting the model to three-dimensional microstructures and conducting comparisons with experimental data to further validate its applicability.

%Bibliography
%\bibliographystyle{unsrt}  
\bibliography{MyLib}  

\appendix
\section{Simulations} \label{appendix:simulations} 
\begin{table}[H]

    \begin{tabular}{p{1.2cm}  p{1.2cm} p{1.2cm} p{1.2cm} p{1.2cm} p{1.2cm} p{1.2cm} p{1.2cm} p{1.2cm} p{1.2cm}}
    \toprule
    \multicolumn{3}{l}{Inputs} &  \multicolumn{4}{l}{Full-field} & \multicolumn{3}{l}{Analytical model}\\
\midrule

       $f_{s}$  & $r_{0}$ & $\sigma/R_{0}$& $\bar{r}_{g}$  & $\bar{r}_{s}$ &  $\bar{r}_{lim}$ & $n_{p}$  & $r_{g}$ & $r_{s}$ &$r_{lim}$ \\
        \hline
 
        1\% & 2.50 & 0.15& 11.0 & 1.86 & 10.6 &252 & 11.2 & 1.66 & 11.13\\
        1\% & 39.3 &0.19 & 46.3 &32.3 & 39.3& 319 &48.0 &32.6  & 39.3\\
        3\% & 2.45  &0.15& 7.37 & 1.67 &6.85 & 148 &7.33 &1.63  &7.17\\
        3\% & 2.50 &0.3& 8.63 &1.93 & 8.15& 105 &7.37  &1.65  & 7.20 \\
        3\% & 4.70&0.3& 8.56 &3.21 &6.48 & 953  &9.55 &3.17 & 8.67\\
        3\% & 4.91& 0.3&9.54 &3.46 & 7.59& 273 &9.75 & 3.32 & 8.77 \\
        3\% & 10.5& 0.3& 14.7 & 6.84 &10.5& 563 & 15.4  & 7.71 & 11.47\\
        3\% &  19.6& 0.3& 24.6 & 14.8 & 19.6 & 366 & 24.6 & 15.9  &19.6 \\
        6\% & 2.43& 0.15& 5.03 &1.54 &4.25 & 378 & 5.85  &1.62 & 5.45 \\
        10\% & 2.36& 0.15& 4.33& 1.64 & 3.49 & 531 & 5.00 &1.58 & 4.33  \\
        15\% & 2.31&0.15 & 3.63 & 1.73 & 2.79 & 802 & 4.48 & 1.57 & 3.58 \\
        20\%& 0.97& 0.4& 2.66 & 0.85 &2.58 & 943 & 2.86  & 0.65 & 2.48 \\
        20\% & 2.25& 0.15&3.43 &1.52 &2.64 & 834 & 4.13 &1.54 & 3.11\\
        20\% & 3.16 &0.13& 4.59& 2.58 & 4.54 & 1739& 5.05& 2.23 & 3.60\\
        20\% & 8.78 &0.15& 9.79& 7.83 & 8.78 & 451 & 10.71 & 7.26 & 8.9\\
        30\% & 1.78 &0.19& 3.16 &1.13 & 2.45 & 969 & 3.32 & 1.21  & 2.39\\
        40\% & 0.80 &0.38& 1.60 &0.62 &2.13  &  709 & 1.26 & 0.54 & 1.58 \\

        \hline

    \end{tabular}
    \caption{Simulation settings, along with simulation outputs, and results obtained from the analytical model of this work. $\sigma$ refers to the standard deviation of the initial distribution.} \end{table}

\end{document}